\documentclass[11pt,preprint]{aastex}

\begin{document}
\newcommand{\lya}{Lyman~$\alpha$}
\newcommand{\lyb}{Lyman~$\beta$}
\newcommand{\degpoint}{\mbox{$^\circ\mskip-7.0mu.\,$}}
\newcommand{\minpoint}{\mbox{$'\mskip-4.7mu.\mskip0.8mu$}}
\newcommand{\secpoint}{\mbox{$''\mskip-7.6mu.\,$}}
\newcommand{\sqdeg}{\mbox{${\rm deg}^2$}}
\newcommand{\squig}{\sim\!\!}
\newcommand{\subsun}{\mbox{$_{\twelvesy\odot}$}}
\newcommand{\et}{{\it et al.}~}
\newcommand{\Rs}{{\cal R}}

\def\ltsima{$\; \buildrel < \over \sim \;$}
\def\simlt{\lower.5ex\hbox{\ltsima}}
\def\gtsima{$\; \buildrel > \over \sim \;$}
\def\simgt{\lower.5ex\hbox{\gtsima}}
\def\propsima{$\; \buildrel \propto \over \sim \;$}
\def\simprop{\lower.5ex\hbox{\propsima}}
\def\arcs{$''~$}
\def\arcm{$'~$}

\title{ESTIMATING THE GALAXY CORRELATION LENGTH $r_0$ FROM
THE NUMBER OF GALAXY PAIRS WITH SIMILAR REDSHIFTS}

\author{\sc Kurt L. Adelberger\altaffilmark{1}}
\affil{Carnegie Observatories, 813 Santa Barbara St., Pasadena, CA 91101}

\altaffiltext{1}{Carnegie Fellow}

\begin{abstract}
We discuss methods that can be used to
estimate the spatial correlation length
$r_0$ of galaxy samples from the observed
number of pairs with similar redshifts.
The standard method is unnecessarily noisy
and can be compromised by errors in the
assumed selection function.  We present
three alternatives, one less noisy, one
that responds differently to systematic errors,
the third insensitive to the selection function,
and quantify their performance by applying
them to a cosmological N-body simulation
and to the Lyman-break survey of galaxies
at redshift $z\sim 3$.  Researchers adopting
the standard method could easily conclude
that the Lyman-break galaxy comoving correlation length
was $r_0 \sim 11h^{-1}$ Mpc, several times larger than
the correct value.  The use of our
proposed methods would make this error impossible,
except in the small sample limit. 
When $N_{\rm gal}\simlt 20$, major errors in estimates
of $r_0$ occur alarmingly often.
\end{abstract}
\newpage

\section{INTRODUCTION}
\label{sec:intro}
This paper was inspired by the work of Daddi et al. (2002, 2004), Blain et al. (2004),
and others who have estimated the spatial clustering strength of a 
galaxy population from the observed positions of a small number of its
members.  Unable to fit a correlation function to the binned numbers
of pair counts at different spatial separations, these authors 
counted the number $n_{\rm obs}$ of galaxy pairs with
redshift separation $|z_1-z_2|<\Delta z$ and compared to the expected
number $n_{\rm exp}$ for an assumed correlation function $\xi(r)$,
which Blain et al. (2004) calculated to be
\begin{equation}
n_{\rm exp} = \frac{N^2}{2\Omega^2}\int_0^{\infty}dz_1 P(z_1)\int_{z_1-\Delta z}^{z_1+\Delta z}dz_2 P(z_2) \int_\Omega d{\bf\Theta_1}\int_\Omega d{\bf\Theta_2} [1+\xi(r_{12})],
\label{eq:blain}
\end{equation}
where $N$ is the number of galaxies with measured redshifts,
$P(z)$ is the survey selection function,\footnote{i.e.,
the redshift distribution that would be observed for an infinitely
large sample in the absence of clustering; our convention
is $\int_0^{\infty}dz\,P(z)=1$.} 
$\Omega$ is the solid
angle of the survey, $r_{12}$ is the comoving distance
between the points specified by $({\bf\Theta_1},z_1)$ and $({\bf\Theta_2},z_2)$,
and ${\bf\Theta}$ is the angular position of a galaxy
within $\Omega$.\footnote{The variable ${\bf\Theta}$ is written in bold-face
because two numbers are required to specify the angular position of an object
on the sky.  If $\alpha$ represents right ascension and $\delta$ represents
declination, $d{\bf\Theta}$ can be interpreted as ${\rm cos}(\delta)d\alpha\,d\delta$.}
They then restricted their attention to a family of correlation functions
$\xi(r)=(r/r_0)^{-1.8}$ that could be specified by a single parameter,
$r_0$, and estimated $r_0$ for their galaxy population by finding the
value that made $n_{\rm exp}=n_{\rm obs}$.  Inspired by 
Poisson statistics, Blain et al. (2004) took as a $1\sigma$ confidence
interval the set of $r_0$ that satisfied
\begin{equation}
n_{\rm obs}-n_{\rm obs}^{1/2} < n_{\rm exp}(r_0) < n_{\rm obs}+n_{\rm obs}^{1/2}.
\label{eq:blainuncertainty}
\end{equation}

The approach can provide useful constraints on $r_0$ when other methods
fail, but the implementation described above is imperfect.
Equation~\ref{eq:blain} is unnecessarily noisy and is more sensitive to
the assumed selection function than to the clustering strength $\bar\xi$;  
equation~\ref{eq:blainuncertainty} almost always underestimates
the true uncertainty in $r_0$.  The goal of this paper is to draw attention to these
shortcomings and to suggest modifications that make the analysis less subject to them.
Section~\ref{sec:selfn} discusses the effect of uncertainties in the selection function,
showing that 
a 20\% error in the assumed width of a Gaussian selection function
can easily change the inferred value of $r_0$ by a factor of 2 or more.
Sections~\ref{sec:angnoise} and~\ref{sec:znoise}
point out two additional sources of noise
in equation~\ref{eq:blain} that are easily removed.
My suggested revisions to the method are put forward
in \S~\ref{sec:alternatives} and tested with a cosmological
$N$-body simulation in \S~\ref{sec:gif}.
Section~\ref{sec:uncertainties} considers the uncertainty
in the best-fit values of $r_0$, showing 
that equation~\ref{eq:blainuncertainty}
is a poor approximation and suggesting a modification
that leads to more realistic error bars.  
The main conclusions are summarized and discussed
in \S~\ref{sec:summary}.  To motivate the discussion, I begin
in \S~\ref{sec:awry} with an example that shows
the standard analysis of redshift pair-counts going badly awry.

\section{A FAULTY ANALYSIS OF LYMAN-BREAK GALAXIES}
\label{sec:awry}
The analyzed sample consists of the 747 Lyman-break galaxies
with apparent magnitude $23.5<{\cal R}<25.5$
in the fields 3c324, b20902, CDFa, CDFb, DSF2237a, DSF2237b, HDF,
Q0201, Q0256, Q0302, Q0933, Q1422, SSA22a, SSA22b, and Westphal
whose spectroscopic redshifts were published by Steidel et al. (2003).
The size of the observed fields varied but was typically $9'\times 9'$.
I calculated the observed number of pairs with comoving radial
separation $Z<20h^{-1}$ Mpc in each field individually.  Summing over all fields,
a total of $n_{\rm obs}=2539$ pairs were found with comoving radial separations in
this range.  Since the Lyman-break technique selects galaxies over
a broad range of redshifts $2.3\simlt z\simlt 3.7$, I approximated
the selection function $P(z)$ as a Gaussian with mean redshift $\mu=3.0$
and standard deviation $\sigma_{\rm sel}=0.4$.  To calculate the
expected number of pairs with $Z<20h^{-1}$ Mpc 
in the $i$th field for a given value
of $r_0$, I inserted this
selection function into equation~\ref{eq:blain}, assumed a correlation
function slope of $\gamma=1.6$, and integrated numerically over the
field's solid angle $\Omega$.  I set the expected total number of pairs $n_{\rm exp}(r_0)$
equal to the sum of the expected number for each individual field.
A value of $r_0=11.08h^{-1}$ Mpc was required for $n_{\rm exp}$ to
equal $n_{\rm obs}$, while $r_0=10.82h^{-1}$ Mpc made 
$n_{\rm exp}=n_{\rm obs}-n^{1/2}_{\rm obs}$ 
and $r_0=11.33h^{-1}$ Mpc made 
$n_{\rm exp}=n_{\rm obs}+n^{1/2}_{\rm obs}$.  I conclude that
the correlation length for Lyman-break galaxies is $r_0=11.1\pm 0.25h^{-1}$ Mpc
at the $1\sigma$ level. 

As noted in the abstract, this estimate of $r_0$ is roughly 20$\sigma$
away from the value of $r_0\simeq 4.0\pm 0.6h^{-1}$ Mpc
measured by Adelberger et al. (2004).  What went wrong?

\section{SOURCES OF ERROR}

\subsection{Selection-function Uncertainties}
\label{sec:selfn}
\begin{figure}
\plotone{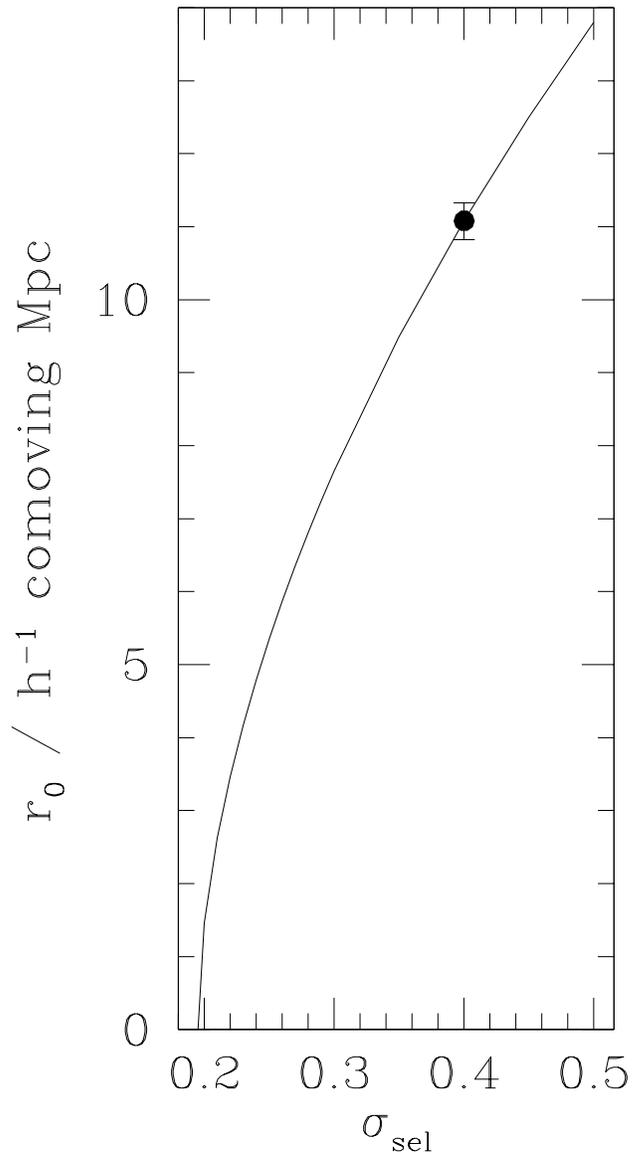}
\caption{
Dependence of the best-fit correlation length in \S~\ref{sec:awry}
on the assumed selection function width $\sigma_{\rm sel}$.  
The point shows the result if $\sigma_{\rm sel}$
is assumed to be $0.4$ exactly.  In fact $\sigma_{\rm sel}$ will
always be somewhat uncertain, and this is one reason that
Poisson error-bars (shown on the point, and
derived from equation~\ref{eq:blainuncertainty}) 
underestimate the true uncertainty.
\label{fig:r0_vs_sigsel}
}
\end{figure}

Most of the error in the previous section's estimate of $r_0$ came from
the inaccurate model of the redshift selection function.
Although it is not always acknowledged in analyses of this sort,
assumptions about the selection function have a critical effect on the results.
Figure~\ref{fig:r0_vs_sigsel} shows that in the example of \S~\ref{sec:awry}
the best-fit value of $r_0$
changes by more than an order of magnitude
as the assumed width of the Gaussian selection function
increases from $\sigma_{\rm sel}=0.2$ to $\sigma_{\rm sel}=0.5$.
If we had adopted the correct width $\sigma_{\rm sel}=0.3$ (Adelberger et al. 2004)
instead of $\sigma_{\rm sel}=0.4$,
we would have found $r_0=7.2$ instead of $r_0=11.1 h^{-1}$ Mpc---significantly closer
to the true value $r_0\sim 4 h^{-1}$ Mpc.

Unfortunately analyses similar to the one in
\S~\ref{sec:awry} are usually attempted
when the sample size is extremely small, too small for
$\sigma_{\rm sel}$ to be determined empirically.  
In this case it is difficult to know which to adopt among the possible values of
$r_0$ suggested by plots similar to Figure~\ref{fig:r0_vs_sigsel}.
Although theoretical arguments may provide a reasonable estimate
of the selection-function shape,  it seems sensible to
reduce as far as possible the dependence of the answer on
the assumed shape. 

Approximating the selection function as a boxcar with half-width $L$, 
equation~\ref{eq:blain} can be rewritten
\begin{equation}
n_{\rm exp}\propto \frac{C}{L}\biggl[1+\bar\xi\biggr]
\label{eq:nexpxibar}
\end{equation}
where $C$ is an uninteresting constant and
$\bar\xi$ is the spatially-averaged correlation function
defined by equation~\ref{eq:blain}.   This form makes it easy to see
why the implied value of $r_0$ can be so strongly affected by
the assumed selection function.  
If the field size $\Omega$
or redshift separation $\Delta z$ is large compared to $r_0$,
as is usually the case, $\bar\xi$ will be significantly less than
unity. The change in $n_{\rm exp}$ that accompanies
a significant change in the correlation strength,
$|dn_{\rm exp}/d{\rm ln}\bar\xi| = C\bar\xi/L$,
will therefore be considerably smaller than the change in
$n_{\rm exp}$ that accompanies significant changes in $L$,
$|dn_{\rm exp}/d{\rm ln}L| = C(1+\bar\xi)/L$,
and minor errors in the assumed selection function
will lead to major errors in the inferred value of $r_0$.
Although these results were derived for a boxcar selection function,
similar results hold for other types.

One way to reduce the method's sensitivity to the selection function
is to design the experiment to maximize $\bar\xi$.
Since $\bar\xi$ increases
as $\Omega$ decreases, experiments with smaller fields-of-view
are less affected by uncertainties in the selection function.
In practice, however, the field-of-view is set by the instrument that is used
and observers are unlikely to want to discard much of their data.
Decreasing $\Delta z$ is a more palatable option, but, owing to
peculiar velocities and to uncertainties in galaxies' measured redshifts,
it cannot be decreased arbitrarily far before genuine pairs
begin to be missed.  $10h^{-1}$ comoving Mpc is a rough lower limit for most
surveys. Unfortunately this limit is large enough to ensure
$\bar\xi\simlt 1$ for likely fields-of-view.

Another way is to use the statistic $K$ (Adelberger et al. 2004)
instead of $n_{\rm obs}$ in the analysis.  This adds noise
but removes the
sensitivity to the selection function almost completely.
The approach is described in more detail below (\S\S~\ref{sec:alternatives}
and~\ref{sec:gif}).

\subsection{Angular distribution of sources}
\label{sec:angnoise}
\begin{figure}
\plotone{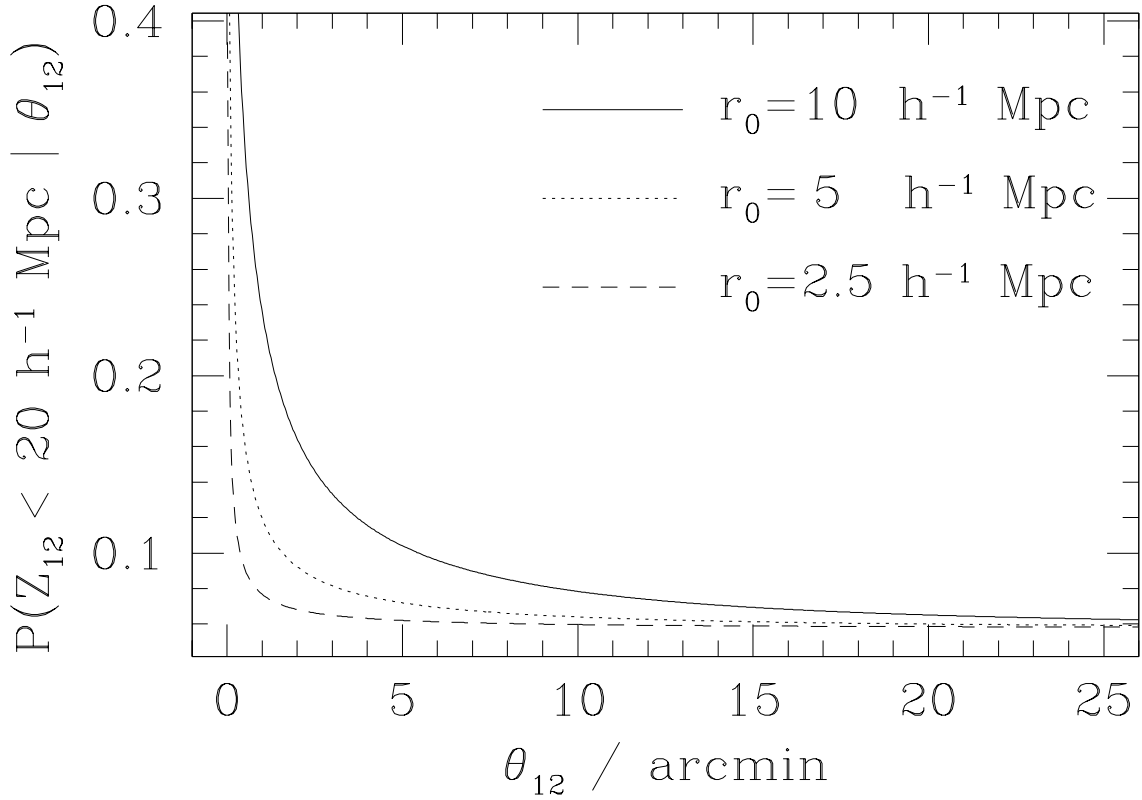}
\caption{
The probability that two galaxies will have comoving radial separation
$Z_{12}<20h^{-1}$ Mpc as a function of the angle $\theta_{12}$ between them.
The actual Lyman-break galaxy selection function (see Figure~\ref{fig:spikealign})
was used in calculating these numbers.  The probability of having
small redshift separations depends at least as much on the galaxies' 
angular separations
as on their correlation length.  This implies that
angular separations should be treated 
carefully when estimating $r_0$ from the number of redshift pairs.
\label{fig:pzlt20}
}
\end{figure}

An additional shortcoming of
equation~\ref{eq:blain} is its assumption that
the sources with measured redshifts have unknown angular positions
that are distributed uniformly across the observed region $\Omega$
(see \S~\ref{sec:case3}).
In fact the angular positions are known (how else were redshifts
measured?) and are probably not uniformly distributed.
Consider, for example, a situation where we obtained images across
a region with radius $r=20'$, but
were able to measure
redshifts for only 2 galaxies.  If these galaxies happened to
have an angular separation of $4''$, they would be likely
to lie at nearly the same redshift even if $r_0$ were small,
while if they had a separation of $40'$ they would be
unlikely to lie at the same redshift even if $r_0$ were
large (Figure~\ref{fig:pzlt20}).
Since the expected
number of close redshift pairs for a known correlation length $r_0$ 
depends on the galaxies' angular separations,
our attempts to infer $r_0$ from the number of pairs will
be improved if we take the galaxies' actual separations into account.
Neglecting this information adds noise to the analysis
and can bias the results if the spectroscopically observed
galaxies were not chosen at random.

\subsection{Redshift distribution of sources}
\label{sec:znoise}
\begin{figure}
\plotone{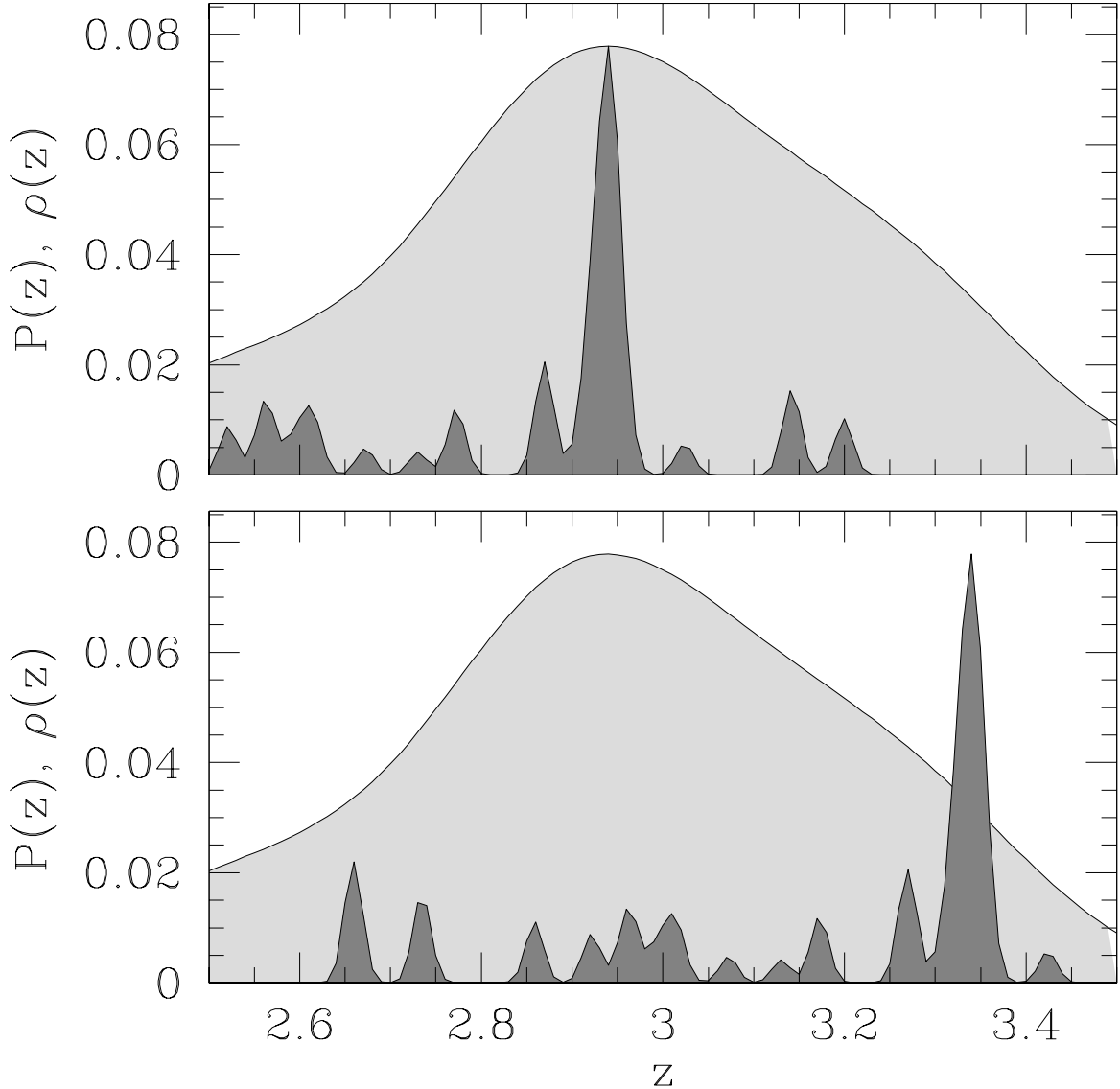}
\caption{
Dependence of the number of redshift pairs on the locations
of galaxy overdensities.  The lighter shaded region in the background
of both panels shows the Lyman-break galaxy selection function $P(z)$.
The darker shaded region shows the galaxy density $\rho(z)$ observed
in the field SSA22a, shifted by $\Delta z=-0.15$ in the top panel
and by $\Delta z=0.25$ in the bottom.  The units on the $y$-axis
are arbitrary.  The galaxy
clustering strength is the same in both panels, but the upper panel
will have roughly $3.5$ times as many pairs with small redshift
separations, on average, since the galaxy overdensity is aligned
with the peak of the redshift histogram and since the number of pairs
is proportional to $\rho^2$.  Estimates of $r_0$ derived solely from
the number of pairs can be led astray by chance alignments or misalignments
of galaxy overdensities with the selection function.  Section~\ref{sec:alternatives}
shows how to remove this unnecessary source of noise;
see the discussion near equation~\ref{eq:nexpgiventhetaz}.
\label{fig:spikealign}
}
\end{figure}

Figure~\ref{fig:spikealign} illustrates another source of noise.  
Suppose we have
found a single galaxy at redshift $z_2$.  How many other galaxies
should we expect to find in the redshift interval
$z_2-\Delta z<z<z_2+\Delta z$ for an assumed value of $r_0$?
The answer depends on the distance between $z_2$ and the peak
of the selection function.  If $z_2$ lies near the peak, we would
expect a large number of pairs even if $r_0$ were small; if $z_2$
lies in the wings we would expect few pairs even if $r_0$ were large.
Since the galaxies in pencil beam surveys 
tend to lie in a small number of prominent spikes in the redshift histogram,
the expected number of redshift pairs is strongly affected by the alignment
or misalignment of the spikes with the peak of the selection function.  
Equation~\ref{eq:blain} is noisier than it needs to be
because it ignores the locations of redshift spikes
when calculating $n_{\rm exp}$.

\section{ALTERNATIVES}
\label{sec:alternatives}
This section suggests alternate approaches that
are less affected by the shortcomings discussed above.
The first two are refinements in the calculation
of $n_{\rm exp}$; the last relies on a slightly
different statistic.  The notation we use is
explained more fully in the appendix.

As shown in the appendix, equation~\ref{eq:blain} 
(in its correctly normalized form, equation~\ref{eq:nexpgivennothing})
gives the number of redshift pairs one should expect to observe
given only the information that $N$ galaxies lie somewhere
in the field of view $\Omega$.
But what if we know angular positions of the sources?
How does this change $n_{\rm exp}$?  If
$P(|Z_{ij}|<\ell\,|\,\theta_{ij})$ is the probability
that a galaxy pair with angular separation $\theta_{ij}$
has comoving radial separation $|Z_{ij}|<\ell$,
then the expected total number of redshift-pairs
should be equal to the sum over all pairs of
$P(|Z_{ij}|<\ell\,|\,\theta_{ij})$:
\begin{equation}
n_{\rm exp} = \sum_{i>j}^{\rm pairs} P(|Z_{ij}|<\ell\, |\, \theta_{ij}).
\label{eq:nexpgiventheta}
\end{equation}
This equation can be evaluated with the help of
equation~\ref{eq:pzgiventheta}.  Using it in place
of equation~\ref{eq:blain} will remove the noise and bias
that arises from the angular positions of the sources.
The estimated correlation length 
of Lyman-break galaxies in our example analysis (\S~\ref{sec:awry}),
reduced from $11.1 h^{-1}$ Mpc to $7.2 h^{-1}$ Mpc by
the adoption of the correct selection function,
is further reduced to $6.4h^{-1}$ Mpc when
equation~\ref{eq:nexpgiventheta} is used instead
of equation~\ref{eq:blain}.
The reduction from $7.2h^{-1}$ to $6.4h^{-1}$ Mpc results partly from
the fact the angular positions of
galaxies with measured redshifts were clumped together
into slitmask-sized regions, not distributed randomly across the field.

How can we incorporate knowledge of the spike redshifts
into the analysis?  Suppose we know that one member of a galaxy
pair with angular separation $\theta_{ij}$ has
the redshift $z_j$.  Then the probability 
$P(|Z_{ij}|<\ell\,|\,z_j\theta_{ij})$
that the
galaxies have radial separation $|Z_{ij}|<\ell$
is given by equation~\ref{eq:pzgiventhetaz}.
The expected total number of
pairs in the sample with redshift separation less than $\ell$
should therefore be equal to the sum of the probabilities for each unique pair,
\begin{equation}
n_{\rm exp} = \frac{1}{2}\sum_{i\neq j}^{\rm pairs} P(|Z_{ij}|<\ell\, |\, z_j\theta_{ij}).
\label{eq:nexpgiventhetaz}
\end{equation}
Using equation~\ref{eq:nexpgiventhetaz} instead of 
equation~\ref{eq:nexpgiventheta} further reduces 
the estimated correlation length (in the example of~\S~\ref{sec:awry})
to $r_0 = 5.7h^{-1}$ Mpc.

Equations~\ref{eq:nexpgiventheta}
and~\ref{eq:nexpgiventhetaz} are as sensitive to errors
in the selection function as equation~\ref{eq:blain}.
This sensitivity can be eliminated almost completely
by using the $K$ statistic of Adelberger et al. (2004)
rather than $n_{\rm obs}$ in the analysis.
Letting $n_{\rm obs}(0,\ell)$
stand for the observed
number of pairs
with comoving radial separation $0\leq |Z_{ij}|<\ell$,
$K$ is the ratio
\begin{equation}
K \equiv \frac{n_{\rm obs}(0,\ell)}{n_{\rm obs}(0,2\ell)}.
\label{eq:defk}
\end{equation}
As long as $n_{\rm obs}(0,2\ell)$ is large enough
that 
\begin{equation}
\Biggl\langle\frac{n_{\rm obs}(0,\ell)}{n_{\rm obs}(0,2\ell)} \Biggr\rangle \simeq \frac{\langle n_{\rm obs}(0,\ell)\rangle}{\langle n_{\rm obs}(0,2\ell) \rangle},
\label{eq:kncondition}
\end{equation}
$K$ will have expectation value
\begin{equation}
\langle K\rangle \simeq \frac{n_{\rm exp}(0,\ell)}{n_{\rm exp}(0,2\ell)}.
\label{eq:expk}
\end{equation}
(In this equation, $n_{\rm exp}(0,\ell)$ can be calculated
with equation~\ref{eq:nexpgiventheta}, equation~\ref{eq:nexpgiventhetaz},
or any number of variants; the value of $K$ will not change
significantly.)
Adelberger et al. (2004) show that the right-hand size of
equation~\ref{eq:expk} is almost entirely independent
of the assumed selection-function width $\sigma_{\rm sel}$ when
$2\ell$ is small compared to $\sigma_{\rm sel}$.  
If we find the value of $r_0$ that makes the
right-hand side of equation~\ref{eq:expk} equal the
right-hand side of equation~\ref{eq:defk}, we will
have an estimate of the correlation length whose value does not depend
on our assumptions about the selection function.\footnote{
Provided the error in the assumed mean redshift is not large enough
to alter significantly the mapping of redshifts and angles onto distances.}
This is our final approach to estimating $r_0$.  Applying
it to the Lyman-break galaxy example of \S~\ref{sec:awry}
leads to an estimate $r_0=4.0h^{-1}$ Mpc that agrees
well with the correlation length reported by
Adelberger et al. (2004).

The discrepancy between the correlation lengths estimated
with equation~\ref{eq:nexpgiventhetaz} and~\ref{eq:expk}
shows that the observed number of pairs with $\ell\leq |Z_{ij}|\leq 2\ell$
is inconsistent with the hypothesis 
$r_0=5.7h^{-1}$ Mpc that seemed (according to
equation~\ref{eq:nexpgiventhetaz}) to account for the number of pairs
with $0\leq |Z_{ij}|\leq \ell$.  This may indicate that
the assumed selection function is incorrect or that
the power-law $\xi(r)=(r/r_0)^{-1.6}$ is a poor approximation
to the correlation function for large separations.
The estimate of $r_0$ will be made more robust against either possibility
by limiting the analysis to pairs with smaller separations,
say $\theta_{ij}<300''$.  In this case the estimated
correlation lengths ($\pm$ standard deviation of the mean from field-to-field
fluctuations)
are 
$5.1\pm 1.1$,
$4.9\pm 0.9$, and
$r_0=4.4\pm 1.1 h^{-1}$ Mpc 
for equations~\ref{eq:nexpgiventheta},~\ref{eq:nexpgiventhetaz}, and~\ref{eq:expk},
respectively, in good agreement with each other and with
the estimate $r_0=4.0\pm 0.6h^{-1}$ Mpc from the angular-clustering
analysis of Adelberger et al. (2004).

The approaches of this section offer two additional benefits.
First, the sum of one-dimensional integrals that they require
is usually simpler to calculate numerically than the
six-dimensional integral required by equation~\ref{eq:blain}.
Second, as we have seen, the form of the equations makes
it easy to omit pairs with undesirable angular separations
from the analysis.

\section{NUMERICAL SIMULATIONS}
\label{sec:gif}
Unimpressed by the heuristic arguments of the previous section,
I tested its recommendations on simulated galaxy surveys
generated 
from the publicly released GIF-$\Lambda$CDM simulation
of structure formation in a cosmology
with $\Omega_M=0.3$, $\Omega_\Lambda=0.7$,
$h=0.7$, $\Gamma=0.21$, $\sigma_8=0.9$.
This gravity-only simulation
contained $256^3$ particles with mass $1.4\times 10^{10} h^{-1} M_\odot$
in a periodic cube of comoving side-length $141.3h^{-1}$ Mpc,
used a softening length of $20 h^{-1}$ comoving kpc, and
was released publicly, along with its halo catalogs, by
Frenk et al. (2000).  Further details
can be found in Jenkins et al. (1998) and Kauffmann et al. (1999).

For the test, I made numerous mock pencil-beam surveys
from the redshift $z=2.32$ catalog of halos with $M>10^{11.2}M_\odot$,
calculated $r_0$ for each mock survey with the approaches
of equations~\ref{eq:blain}, \ref{eq:nexpgiventheta},
\ref{eq:nexpgiventhetaz}, and~\ref{eq:expk},
then tabulated and compared the results.
To generate a single mock pencil-beam survey from the cubical
simulation, I concatenated numerous randomly selected volumes
of size $13\times 13\times 141.3h^{-3}$ Mpc$^3$
into a long parallepiped with dimension
$13\times 13\times 1700 h^{-3}$ Mpc$^3$.
After converting the comoving coordinates
of each halo in the volume into redshift and angle
(for $\Omega_M=0.3$, $\Omega_\Lambda=0.7$, with
$1700h^{-1}$ Mpc the redshift depth),
I applied various selection effects to produce
one mock pencil beam survey.  
Numerous additional mock surveys, each generated in the same way,
were used in the analysis.
The mock surveys are clearly not
exact reproductions of the actual universe.  They are
discontinuous every $141.3h^{-1}$ Mpc, do not
include any evolution in structure from the back to
the front of the volume,
and have an incorrect power-spectrum
on very large ($\simgt 141h^{-1}$ Mpc) scales because they
were extracted from a single $141.3h^{-1}$ Mpc cube.
However,
the methods of \S~\ref{sec:alternatives} work for
objects with any spatial distribution, as long
as the correlation function is sharply peaked,
and the simulated surveys are similar enough to
actual redshift surveys to provide a meaningful
preview of how equations~\ref{eq:blain}, \ref{eq:nexpgiventheta},
\ref{eq:nexpgiventhetaz}, and~\ref{eq:expk},
will behave in realistic situations.

The results are summarized in Figure~\ref{fig:gif_summary}.  All panels
are for a simulated survey with a $10'\times 10'$ field of view.
The correlation function slope was fixed to $\gamma=1.6$ and
$\ell=20h^{-1}$ Mpc was taken as the maximum pair separation.
The panel on the upper left shows the distribution of
estimated $r_0$ from the four techniques when
the pencil beam surveys included $N_{\rm gal}=200$ galaxies each
and had a Gaussian selection function
with mean $\mu_z=2.2$ and r.m.s. $\sigma_z=0.35$.
I used the correct selection function 
in calculating $r_0$ for the idealized case of
this panel, even though normally $r_0$
will be calculated from an assumed  selection function
that is at least somewhat incorrect.
This panel provides a reference against which the others
can be judged.  

The catalogs for the other panels 
were constructed in the same way, except as noted below.
The middle left panel  shows
the effect of lowering $N_{\rm gal}$ from 200 to 20.  The noise
in $r_0$ increases significantly with catalogs so
small.  The estimates become biased 
because the dependence of $r_0$ on the number of pairs $n$
is no longer approximately linear  over the plausible range of $n$.
Although no approach performs particularly well, 
the method of equation~\ref{eq:expk}
is essentially unusable.  This is because
random fluctuations in pair counts
often make $n_{\rm obs}(0,\ell)=n_{\rm obs}(0,2\ell)$,
and the equivalent relationship for $n_{\rm exp}$
requires $r_0\to\infty$.  (More formally, it
is because equation~\ref{eq:kncondition} is
no longer a good approximation.)

For the bottom left panel, the simulated galaxies' angular positions
were concentrated towards the 
center of the field rather than being random:
each galaxy's selection probability was multiplied
by a Gaussian with $\sigma=70''$ centered in the middle
of the field, causing
90\% of the galaxies in a typical catalog to fall 
within a region of diameter $5'$ 
inside the larger $10'\times 10'$ field.
This was intended to mimic the sort of selection effect
than can appear in multislit spectroscopic surveys.
In this case equation~\ref{eq:blain} leads to biased
results, since it makes incorrect assumptions
about the galaxies' angular positions, while the
three approaches of
\S~\ref{sec:alternatives} are nearly unaffected.

The upper right panel shows the what happens
to the inferred value of $r_0$ if the 
expected pair counts are calculated under the
erroneous assumption that the
selection-function width is $\sigma_z=0.5$.
(In all panels its true value is $\sigma_z=0.35$.)
Equations~\ref{eq:blain} and~\ref{eq:nexpgiventheta}
fare the worst, producing estimates of
$r_0$ that are two high by a factor of two.   
Equation~\ref{eq:nexpgiventhetaz}
leads to smaller errors, but only because $\sigma_z$ was overestimated;
for underestimates it performs worse.
Only equation~\ref{eq:expk} yields unbiased results.

The middle right panel shows what happens when the 
assumed selection function has the correct width
$\sigma_z=0.35$ but the incorrect
mean, $\mu_z=2.8$, 
instead of the true value $\mu_z=2.2$.
Equations~\ref{eq:blain} and~\ref{eq:nexpgiventheta}
produce underestimates of $r_0$, because the
selection function is assumed to be narrower in comoving units
than it actually is.
Equation~\ref{eq:nexpgiventhetaz} produces an overestimate,
doing more harm than good in its mangled attempts to 
compensate for the alignment
of the selection function with redshift spikes.
Equation~\ref{eq:expk} remains satisfactory.

The bottom right panel shows a worst case scenario,
which may be closer than any other panel to
actual cases found in the literature.
The sample size is $N_{\rm gal}=20$, the data are
subject to angular selection effects (modeled by a two dimensional
Gaussian distribution that has $90$\%
of sources within a region of diameter $7.2'$),
and the pair counts have been analyzed under the assumption
that $\mu_z=2.2$, $\sigma_z=0.5$ even though
the true selection function has $\mu_z=2.2$, $\sigma_z=0.35$.
The results here are so uncertain and biased as to be useless.
Estimates $r_0>10h^{-1}$ Mpc appear alarmingly often,
compensated only by the common occurence of $r_0=0$.
Adopting equation~\ref{eq:nexpgiventheta} or~\ref{eq:nexpgiventhetaz}
helps reduce the noise, but none of the approaches are likely
to add significantly to the observer's prior knowledge of $r_0$.

\section{UNCERTAINTIES}
\label{sec:uncertainties}
Equation~\ref{eq:blainuncertainty} produces
a reasonable estimate of the uncertainty in the
simulation results if the ``Poisson'' uncertainty
$n_{\rm obs}^{1/2}$ is replaced with
the true uncertainty ${\rm Var}(n_{\rm obs})^{1/2}$,
where ${\rm Var}(n)$ is short-hand for the variance
of $n$.
As Figure~\ref{fig:varn_vs_n} shows, the two can differ
significantly; the
clustering of galaxies drives the variance
in pair counts far above the Poisson value ${\rm Var}(n)=n$.

The variance of $n_{\rm obs}$ is easy
to estimate for the ensemble of simulated surveys.  
As long as random errors dominate over cosmic variance,
it can be estimated
in real life by splitting a survey into many smaller subsamples,
calculating the dispersion in $n_{\rm obs}$ among the
subsamples, measuring how the dispersion changes with
subsample size, and extrapolating to the full sample size.
Sample results are shown in Figure~\ref{fig:varn_vs_n}.  For the Lyman-break survey,
this approach 
leads to an estimated
$1\sigma$ uncertainty in $r_0$ of $\sim 1.3h^{-1}$ Mpc,
which agrees well with the value
$\sigma_{\rm ftf}/N^{1/2}\simeq 1.1h^{-1}$ Mpc
implied by the field-to-field fluctuations $\sigma_{\rm ftf}$
in the estimated value of $r_0$ from the $N=15$ individual
survey fields.

The preceding discussion applies to values of
$r_0$ estimated from 
equations~\ref{eq:blain}, \ref{eq:nexpgiventheta},
and~\ref{eq:nexpgiventhetaz}, since in these
cases $r_0$ is estimated by setting $n_{\rm obs}=n_{\rm exp}$.
The uncertainties are slightly more difficult to
estimate in the case of equation~\ref{eq:expk}.
One approach, in this case, is to estimate
the dispersion in $r_0$, not $n_{\rm obs}$, among the
subsamples, and extrapolate this to the full sample size.

These procedures do not work well for small samples,
but neither do the methods for estimating $r_0$ itself.  I discuss
this further in the summary section.

\section{SUMMARY}
\label{sec:summary}
\begin{figure}
\plotone{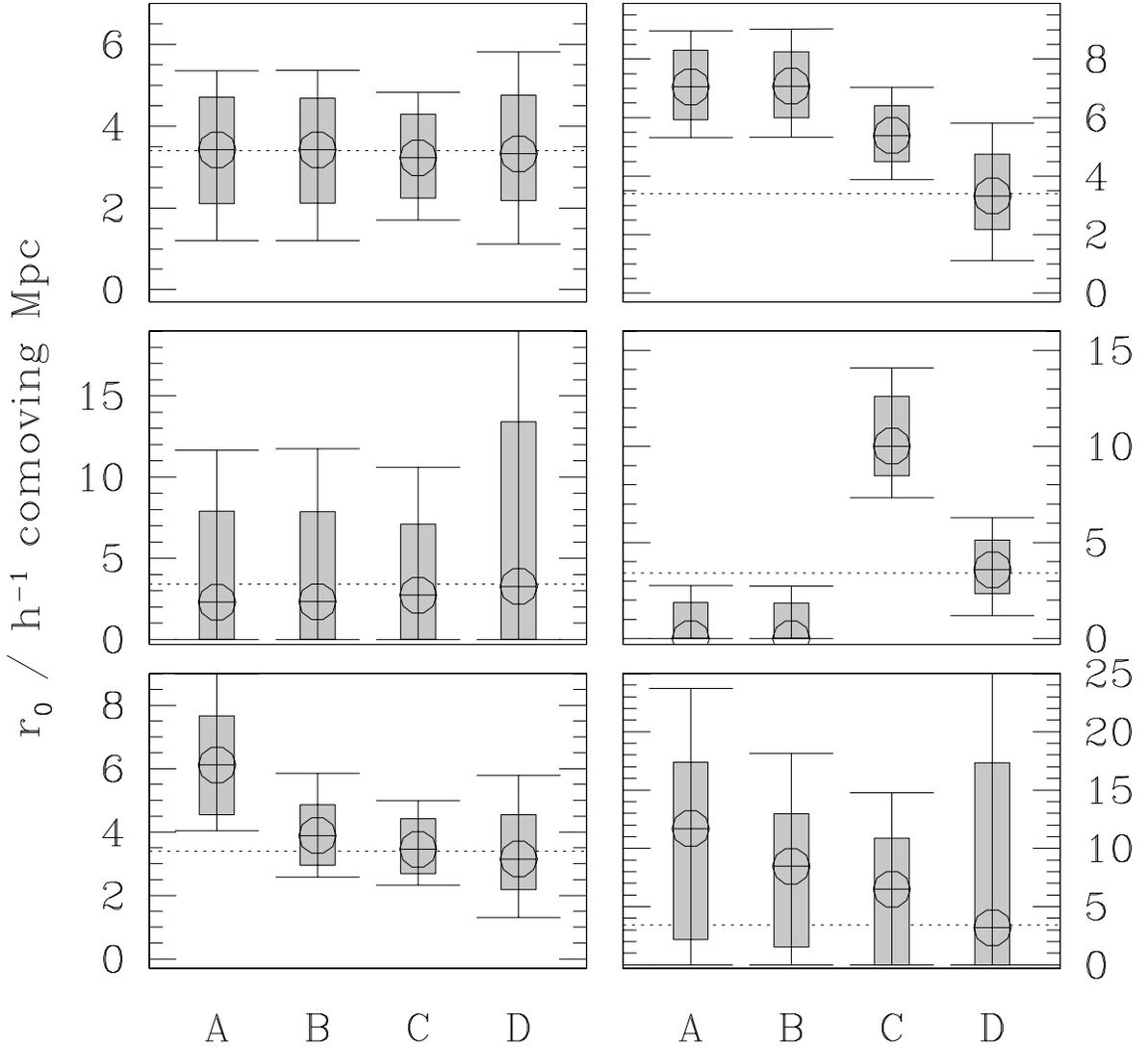}
\caption{
Performance of the 4 methods on simulated galaxy surveys.
Each panel shows the results for surveys generated
with a single parameter combination (\S~\ref{sec:gif}).
Points labeled $A$, $B$, $C$, and $D$ summarize the
results for the methods that use equations~\ref{eq:blain},
\ref{eq:nexpgiventheta}, \ref{eq:nexpgiventhetaz},
and~\ref{eq:expk}, respectively.  
(Equation~\ref{eq:blain} was actually used
in its correctly normalized form, equation~\ref{eq:nexpgivennothing}.)
All fits assumed a correlation function slope $\gamma=1.6$
and adopted $\ell=20h^{-1}$Mpc as the maximum pair separation.
The circle marks the
median estimate of $r_0$; the estimates fell within
the shaded region for 68\% of the simulated surveys,
and within the error bars for 90\%.
The horizontal dashed line shows the true value of $r_0$,
calculated by counting the number of pairs as a function of
separation for all halos in the GIF catalog,
then fitting a power-law correlation function to the result.
The upper left panel is for a survey with $N=200$ galaxies
in a single $10'\times 10'$ field where the
true selection function (a Gaussian with $\mu_z=2.2$, $\sigma_z=0.35$)
is used in the analysis.  Survey parameters are varied
in other panels.  Middle left: $N=20$.  Bottom left:  
spectroscopic selection effects concentrate the survey
galaxies near the center of the field.
Upper right: a selection function with incorrect width 
($\sigma_z=0.5$)
is used in the analysis.  Middle right: a selection function
with incorrect mean ($\mu_z=2.8$) is used in the analysis.
Bottom right panel: $N=20$, angular selection effects,
incorrect selection function used in the analysis.
Further details are given in \S~\ref{sec:gif}.
\label{fig:gif_summary}
}
\end{figure}

\begin{figure}
\plotone{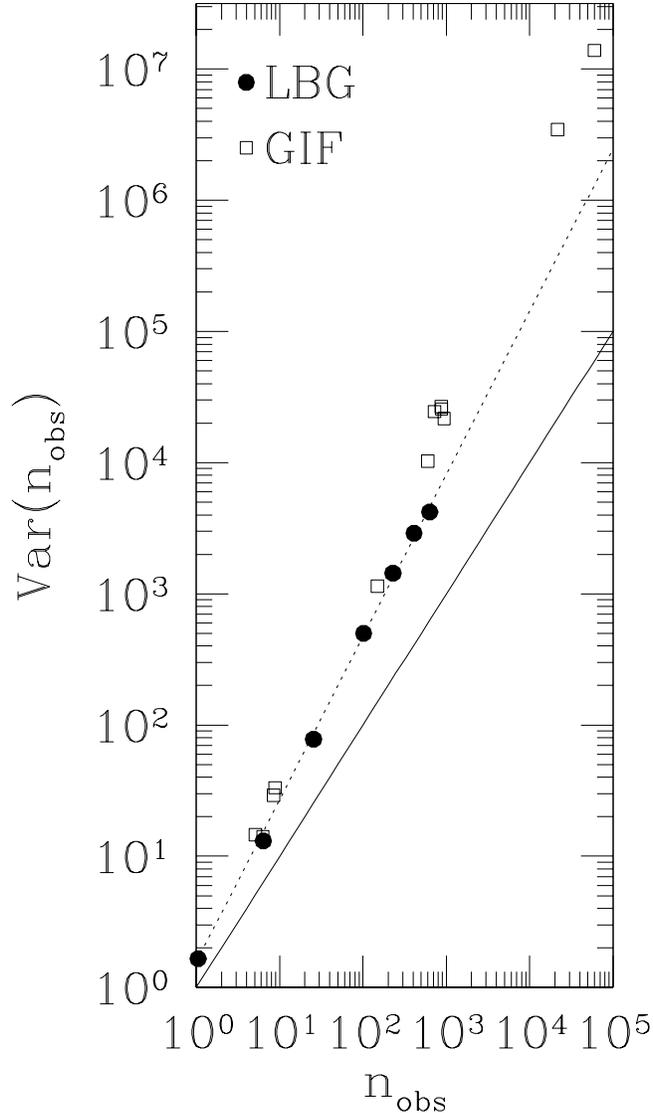}
\caption{
Dependence of the variance of pair-counts on the number of pairs.
Results are shown for Lyman-break galaxies (filled circles) 
and for halos in the GIF simulation (open squares).
To estimate the dependence for LBGs,
we created numerous subsamples with different mean
numbers of galaxies by eliminating a random fraction
of galaxies from the
actual LBG catalogs described in \S~\ref{sec:awry}.
Seven sets of subsamples were created, with the eliminated
fraction $f=0.98$, 0.95, 0.9, 0.8, 0.7, 0.6, and 0.5.
The point with $n_{\rm obs}\simeq 10^{2.8}$, ${\rm Var}(n_{\rm obs})\simeq 10^{3.6}$
shows the mean and variance of the number of galaxy pairs with 
radial separation $Z_{12}<20h^{-1}$ Mpc in the subsamples with $f=0.5$.
The other points are defined similarly.  
Each GIF point (open square) shows $\langle n\rangle$ and the variance in $n$
for the ensemble of simulated pencil-beam surveys 
created for a single set of mock survey parameters (\S~\ref{sec:gif}).
These parameter sets include but are not limited to the ones
shown in Figure~\ref{fig:gif_summary}.
The ``Poisson'' approximation
${\rm Var}(n)=n$ (solid line) is poor for all values of $n_{\rm obs}$.
A better approximation, for the LBG survey, is ${\rm Var}(n)=1.56n^{1.24}$
(dotted line).  Different relationships will hold for
different surveys, as the GIF results show, and this
relationship should not assumed in other situations.
A sensible way to estimate ${\rm Var}(n)$ for other
surveys is to create random subsamples like these, fit a function
to the curve of ${\rm Var}(n)$ vs. $n$, and extrapolate to the observed number
of pairs.
\label{fig:varn_vs_n}
}
\end{figure}

This paper analyzed a method that has recently been used
to estimate
the spatial clustering strength of small galaxy samples.
The method is imperfect.  The
estimate of $r_0$ (a) depends sensitively on the assumed selection
function (Figure~\ref{fig:r0_vs_sigsel}), (b) will be biased
if the galaxies are not distributed approximated uniformly across
the field (Figure~\ref{fig:gif_summary}), 
and (c) is strongly affected by the positions of
galaxy overdensities relative to the peak of the selection function
(Figure~\ref{fig:spikealign}).  

I suggested three ways
to mitigate these problems and tested my suggestions
on simulated galaxy surveys and on the Lyman-break survey.
Figure~\ref{fig:gif_summary} provides a useful overview
of the results.
When there are no systematic errors, equation~\ref{eq:nexpgiventhetaz}
produces the best estimates of $r_0$ and
equation~\ref{eq:expk} the worst.  Equation~\ref{eq:expk}
is robust against systematic errors, however, and continues
to produce reasonable estimates in the presence of
systematic effects that render the other approaches useless.
Since the approaches respond differently to systematic and
random errors, a sensible strategy
is to estimate $r_0$ with all of them\footnote{except
equation~\ref{eq:blain}; as far as I can tell, there is no situation
where its performance is the best among the alternatives}
and look
for consistency among the results.  

The sample analysis of the Lyman-break survey helps
illustrate the paper's main points.
An initial estimate of $r_0\sim 11h^{-1}$ Mpc
from equation~\ref{eq:blain} disagreed badly with the
estimate $r_0\sim 4h^{-1}$ Mpc from the robust equation~\ref{eq:expk},
suggesting that the initial analysis must have
had large systematic errors.  The largest systematic error
came from inaccuracies in the assumed selection
function.  Replacing it with a better model reduced
the estimated values of $r_0$ to
$7.2$, $6.4$, $5.7$, and $4.0h^{-1}$ Mpc from
equations~\ref{eq:blain}, \ref{eq:nexpgiventheta},
\ref{eq:nexpgiventhetaz}, and~\ref{eq:expk}, respectively.
The differences were still not negligible compared to
the random uncertainties (\S~\ref{sec:uncertainties}).
The high value from equation~\ref{eq:blain} was due
to artificial angular clustering of galaxies imposed
by the survey's spectroscopic selection criteria.
It alone among the estimators does not correct for this.
The remaining systematic problems are not easy to trace.
They could result from residual errors in the
selection function or from changes in the correlation function
slope at large separations.  In any case,
since the effect of systematic errors is minimized when they
are small compared to the signal, I maximized
the signal by limiting
the analysis to angular pairs with smaller separations.
As equation~\ref{eq:nexpxibar} shows,
the number of pairs with large angular separations
is more sensitive to low level systematics than
to the clustering
strength $\bar\xi$. 
Restricting the analysis to pairs with angular separation
$\theta_{ij}<300''$, I obtained the estimates
$r_0=5.1h^{-1}$, $4.9h^{-1}$, $4.4h^{-1}$ Mpc from
equations~\ref{eq:nexpgiventheta}, \ref{eq:nexpgiventhetaz},
and~\ref{eq:expk}.
Since the random uncertainty is $\sim 1h^{-1}$ Mpc (\S~\ref{sec:uncertainties}),
these estimates agree well with each other and with
the value $r_0=4.0\pm 0.6h^{-1}$ Mpc favored by the angular-clustering
analysis of Adelberger et al. (2004).

\begin{figure}
\plotone{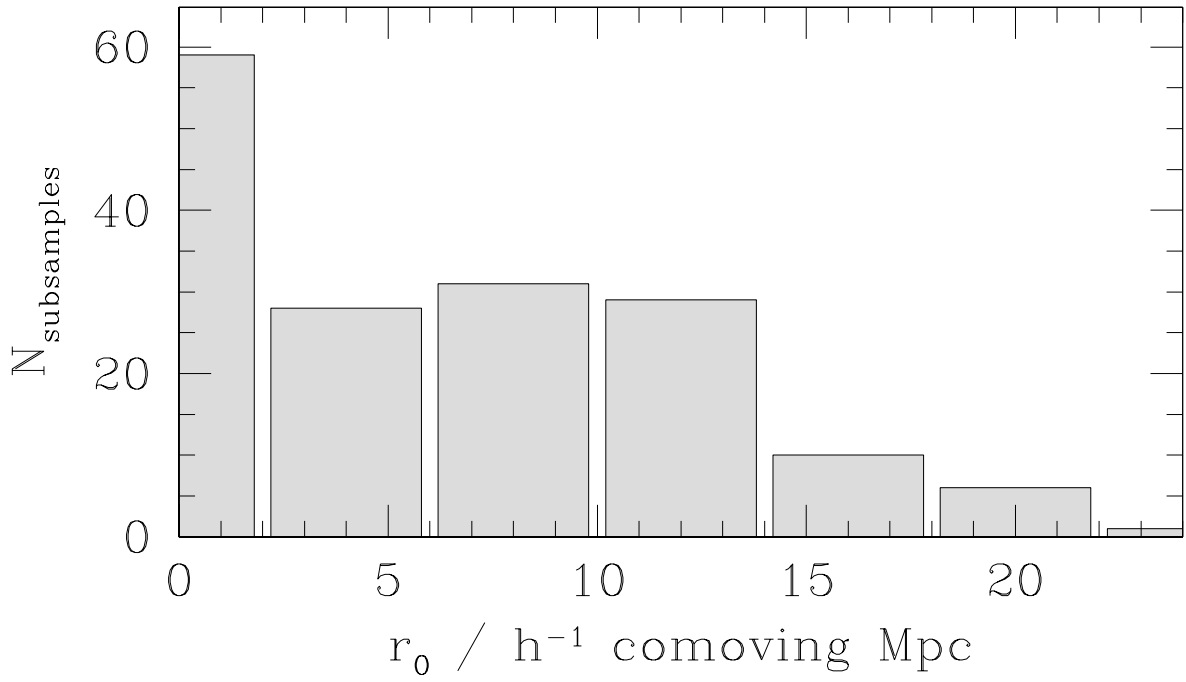}
\caption{
Distribution of $r_0$ for 10-galaxy LBG subsamples
extracted at random from the 170-object Westphal catalog
of Steidel et al. (2003).  Correlation lengths were estimated
with the approach of equation~\ref{eq:blain}.
\label{fig:westconfint}
}
\end{figure}

This paper provides some support for the common prejudice against
estimates of $r_0$ derived from small galaxy samples.
The middle left panel of Figure~\ref{fig:gif_summary}
shows how large the random uncertainties are
for a simulated sample of $N=20$ galaxies with true correlation
length $r_0=3.5h^{-1}$ Mpc in a $10'\times 10'$ pencil-beam survey.
Figure~\ref{fig:westconfint} may make the point more forcefully.
I extracted numerous random subsamples of 10 galaxies from the 
170-object Lyman-break galaxy catalog in the Westphal field
(Steidel et al. 2003), calculated $r_0$ for each subsample
with equation~\ref{eq:blain} using the true LBG selection function, and
tabulated the results.  The spread in estimated $r_0$ is enormous.

In realistic situations, uncertainty in the assumed selection function 
is likely to be the worst source of systematic error.
A skeptic might point out that this uncertainty will probably
only be large in the small sample limit, where none of the
approaches work well, and that my suggested alternatives
are not much of an improvement when the uncertainty in the
selection function is small (see, e.g., the upper left panel
of Figure~\ref{fig:gif_summary}).  This is true to a point, 
but it would be foolish to reject the
$\sim 30$\% reduction in random uncertainty that
equation~\ref{eq:nexpgiventhetaz} provides relative to
equation~\ref{eq:blain}.  According to Figure~\ref{fig:varn_vs_n},
a $\sim 30$\% decrease in the uncertainty in $r_0$
for the LBG sample requires a $\sim 40$\% increase in the
number of galaxies.  Using equation~\ref{eq:nexpgiventhetaz}
instead of~\ref{eq:blain} in the analysis is surely easier
than requesting, obtaining, and reducing $40$\% more data. 
The methods of \S~\ref{sec:alternatives} are far from perfect,
but they improve significantly on their predecessor.

\bigskip
\bigskip
I would like to 
thank the Florida Airport cafe in La Serena
for its hospitality while
the first draft of this paper was being written.
My collaborators in the Lyman-break survey encouraged me
to share the analysis with a wider audience.  
This work was supported by a fellowship from the Carnegie
Institute of Washington.

\appendix
\section{EXPECTED PAIR COUNTS FOR POWER-LAW CORRELATION FUNCTIONS}
We derive three simple results needed in the text.  In each case,
the notation $P(AB|C)$ stands for the probability
that $A$ and $B$ are both true if we know that $C$ is true.
According to this notation, 
$P(z_1{\bf\Theta_1}|z_2{\bf\Theta_2})$ is the
probability of finding a galaxy at
redshift $z_1$ and angular position ${\bf\Theta_1}$
if we know that there is a galaxy at position $z_2, {\bf\Theta_2}$,
and $P(z_1{\bf\Theta_1})$ is the probability
of finding a galaxy at the first position if we know nothing about
the positions of other galaxies.  We assume that the reduced two-point galaxy
correlation function, $\xi$, is an isotropic power-law, 
$\xi(r)=(r/r_0)^{-\gamma}$, which implies that
\begin{equation}
P(z_1{\bf\Theta_1}|z_2{\bf\Theta_2})=P(z_1{\bf\Theta_1})[1+(r_{12}/r_0)^{-\gamma}]
\end{equation}
where $r_{12}$ is the distance between the points specified by
$z_1,{\bf\Theta_1}$ and $z_2,{\bf\Theta_2}$.
Since the survey selection function
does not depend on sky position, $P(z_1{\bf\Theta_1})=P(z_1)/\Omega$
where $\Omega$ is the survey's solid angle and
$P(z_1)$ is the expected redshift distribution for a single
object in our survey.  
We adopt the shorthand $z_{12}\equiv z_1-z_2$ and
$\theta_{12} \equiv |{\bf\Theta_1}-{\bf\Theta_2}|$,
and use the capitalized variable $Z_{12}$ to indicate
comoving distance between redshifts $z_1$ and $z_2$.

\subsection{Case 1:}
If we know that a galaxy with position $z_2$, ${\bf\Theta_2}$
has a neighbor at angular position ${\bf\Theta_1}$, what is the 
probability that the neighbor has redshift $z_1$?
In our notation, we are asking for $P(z_1|z_2{\bf \Theta_1\Theta_2})$,
which can be derived from the correlation function with elementary probability
identities:
\begin{eqnarray}
P(z_1|z_2{\bf \Theta_1\Theta_2}) &=& \frac{P(z_1{\bf\Theta_1}|z_2{\bf\Theta_2})}{\int_0^\infty dz_1'\,P(z_1'{\bf\Theta_1}|z_2{\bf\Theta_2})} \\
 &\simeq& \frac{P(z_1) [1+\xi(r_{12})]}{1+a(r_0,\gamma,\theta_{12},z_2)P(z_2)}.
\label{eq:pz1givenz2theta}
\end{eqnarray}
The second equality assumes that the selection function is
independent of angular position ${\bf\Theta}$ and is roughly constant
over the small radial separations where $\xi$ is significantly larger
than 0.  It also assumes that $f$ and $g$ (defined in the following sentence)
do not change significantly over the same small radial separations.
For clarity we adopt the shorthand
\begin{equation}
a(r_0,\gamma,\theta,z)\equiv r_0^\gamma[f(z)\theta]^{1-\gamma}g^{-1}(z)\beta(\gamma)
\end{equation}
where 
$g(z)\equiv c/H(z)$ is the change in comoving distance with redshift,
$f(z)\equiv (1+z)D_A(z)$ is the change in comoving distance with angle,
$D_A(z)$ is the angular diameter distance, 
$\beta(\gamma)\equiv B[1/2,(\gamma-1)/2]$,
and $B$ is the beta function in the convention of Press et al. (1992).

The probability that the comoving distance $|Z_{12}|$ between $z_1$ and $z_2$ will
be less than $\ell$ can be derived by integrating equation~\ref{eq:pz1givenz2theta}
over the appropriate range of $z_1$:
\begin{eqnarray}
P(|Z_{12}|<\ell\, |\, z_2\theta_{12}) &=& \frac{\int_{z_2-\ell/g}^{z_2+\ell/g}dz_1\,P(z_1{\bf\Theta_1}|z_2{\bf\Theta_2})}{\int_0^\infty dz_1'\,P(z_1'{\bf\Theta_1}|z_2{\bf\Theta_2})} \\
                                      &\simeq& \frac{P(z_2) [2\ell g^{-1}(z_2)+a(r_0,\gamma,\theta_{12},z_2){\cal I}(\gamma,{\ell},\theta_{12},z_2)]}{1+a(r_0,\gamma,\theta_{12},z_2)P(z_2)}
\label{eq:pzgiventhetaz}
\end{eqnarray}
where ${\cal I}$ is related to the incomplete beta function $I_x$ 
of Press et al. (1992) through
\begin{equation}
{\cal I}(\gamma,{\ell},\theta,z) \equiv I_x[1/2,(\gamma-1)/2]
\end{equation}
with
\begin{equation}
x\equiv \frac{\ell^2}{\ell^2+[f(z)\theta]^2}.
\end{equation}

\subsection{Case 2:}
What is the probability that the galaxy pair with known angular separation
$\theta_{12}$ has comoving redshift separation $|Z_{12}|<\ell$?
The probability that a pair with angular separation $\theta_{12}$ will have redshift
separation $z_{12}$ is
\begin{equation}
P(z_{12}|\theta_{12}) = \int_0^\infty dz_2\, P(z_2|\theta_{12}) P(z_{12}|z_2\theta_{12}),
\end{equation}
which implies that the pair will have comoving radial separation $|Z_{12}|$
less than $\ell$ with probability
\begin{eqnarray}
P(|Z_{12}|<{\ell}\, |\, \theta_{12}) &=& \int_0^\infty dz_2 P(z_2) \frac{\int_{z_2-\ell/g}^{z_2+\ell/g}dz_1\,P(z_1{\bf\Theta_1}|z_2{\bf\Theta_2})}{\int_0^\infty dz_1'\,P(z_1'{\bf\Theta_1}|z_2{\bf\Theta_2})} \\
                                     &\simeq& \int_0^\infty dz P^2(z)\frac{2{\ell}g^{-1}(z)+a(r_0,\gamma,\theta_{12},z){\cal I}(\gamma,{\ell},\theta_{12},z)}{1+a(r_0,\gamma,\theta_{12},z)P(z)}.
\label{eq:pzgiventheta}
\end{eqnarray}

\subsection{Case 3:}
\label{sec:case3}
What is the expected number of pairs with $|Z_{ij}|<\ell$ if we know only 
that $N$ galaxies lie somewhere in the solid angle
angle $\Omega$?  If $I_i$ represents the proposition that galaxy $i$ lies
within the surveyed solid angle $\Omega$, the expected number 
of pairs will depend on $\int_{-\ell}^{\ell} dZ_{12} P(Z_{12}|I_1I_2)$,
the probability that a randomly selected pair in the survey 
has comoving redshift separation
less than $\ell$.
The conditional probability can be rewritten as
\begin{equation}
P(Z_{12}|I_1I_2) = \frac{P(Z_{12}I_1I_2)}{\int_{-\infty}^{\infty} dZ_{12} P(Z_{12}I_1I_2)}
\label{eq:pz12giveni1i2}
\end{equation}
and the unconditional probability can be expanded to
\begin{equation}
P(Z_{12}I_1I_2) = \int d{\bf\Theta_1}d{\bf\Theta_2}dz_2\, P(Z_{12}{\bf\Theta_1}{\bf\Theta_2}I_1I_2z_2)
\label{eq:pz12i1i2}
\end{equation}
where the integrals in equation~\ref{eq:pz12i1i2}  extend over all space.
If ${\bf\Theta_1}$ is not within $\Omega$,
$P(Z_{12}{\bf\Theta_1}{\bf\Theta_2}I_1I_2z_2)$ will be equal to 0.
If ${\bf\Theta_2}$ is within $\Omega$,
$P(Z_{12}{\bf\Theta_1}{\bf\Theta_2}I_1I_2z_2)$ will be equal to
$P(Z_{12}{\bf\Theta_1}{\bf\Theta_2}I_2z_2)$ for the same reason that
the probability of being in the Louvre and in France is equal to 
the probability of being in the Louvre.
Since the same arguments apply to ${\bf\Theta_2}$ and $I_2$,
equation~\ref{eq:pz12i1i2} can be simplified by omitting
$I_1$ and $I_2$ from the right-hand side and restricting
the angular integrals to the region $\Omega$.  After expanding
the integrand with the identify $P(AB)=P(A|B)P(B)$, equation~\ref{eq:pz12i1i2} becomes
\begin{equation}
P(Z_{12}I_1I_2) = \int_\Omega d{\bf\Theta_1}d{\bf\Theta_2}\int_0^{\infty} dz_2 P(Z_{12}{\bf\Theta_1}|z_2{\bf\Theta_2}) P(z_2{\bf\Theta_2}).
\label{eq:pz12i1i2b}
\end{equation}
The expected number of pairs with $|Z_{12}|<\ell$ is equal to the number of unique pairs multiplied
by the probability that a random pair has $|Z_{12}|<\ell$.  Substituting
equation~\ref{eq:pz12i1i2b} into equation~\ref{eq:pz12giveni1i2} and
integrating over $Z_{12}$, one finds
\begin{eqnarray}
n_{\rm exp} &=& \frac{N(N-1)}{2} \frac{\int_\Omega d{\bf\Theta_1}d{\bf\Theta_2}\int_0^{\infty} dz_2\,P(z_2)\int_{z_2-\ell/g}^{z_2+\ell/g}dz_1\,P(z_1)[1+\xi(r_{12})]}{\int_\Omega d{\bf\Theta_1}d{\bf\Theta_2}\int_0^{\infty} dz_2\,P(z_2)\int_{0}^{\infty}dz_1\,P(z_1)[1+\xi(r_{12})]} \\
            &=& \frac{N(N-1)}{2} \frac{\int_\Omega d{\bf\Theta_1}d{\bf\Theta_2}\int_0^{\infty} dz\,P^2(z) [2\ell g^{-1}(z)+a{\cal I}]/\Omega^2}{1+r_0^\gamma \beta(\gamma)\int_\Omega d{\bf\Theta_1}d{\bf\Theta_2}\theta_{12}^{1-\gamma}\int_0^{\infty} dz\,P^2(z)f^{1-\gamma}(z) g^{-1}(z)/\Omega^2}
\label{eq:nexpgivennothing}
\end{eqnarray}
which recovers equation~\ref{eq:blain}, aside from the latter equation's
imprecise normalization.

\newpage

\newpage

\newpage

\newpage

\newpage

\newpage

\end{document}